\documentclass[final,5p,times,twocolumn]{elsarticle}

\usepackage{lineno}
\usepackage{graphicx}
\usepackage{amssymb}
\usepackage{pdflscape}
\usepackage[english]{babel}
\usepackage[dvipsnames,svgnames,x11names]{xcolor}
\usepackage{booktabs,tabularx}
\usepackage{multirow}
\usepackage{hyperref}
\usepackage{url}            
\usepackage{booktabs}       
\usepackage{amsfonts}       
\usepackage{nicefrac}       
\usepackage{microtype}      
\usepackage{amsmath}
\usepackage{commath}
\usepackage[english]{babel}
\usepackage[ruled]{algorithm2e}
\usepackage{caption}
\usepackage{subcaption}
\usepackage{lipsum}
\usepackage{graphicx}
\usepackage{makecell}
\usepackage{longtable}
\usepackage{enumitem}
\usepackage[printonlyused]{acronym}
\usepackage[section]{placeins}
\usepackage[compact]{titlesec} 
\usepackage{setspace}
\SetEndCharOfAlgoLine{}
\usepackage[leftmargin=6em,rightmargin=12em,indentfirst=false]{quoting}
\usepackage{siunitx}

\usepackage[final]{changes}
\usepackage{float}

\usepackage{lineno}
\usepackage{tikz}
\usepackage[export]{adjustbox}

\usepackage{graphicx}
\usepackage[utf8]{inputenc}
\usepackage[export]{adjustbox}
\usepackage{wrapfig}
\usepackage{dcolumn}

\makeatletter
\newcolumntype{T}[3]{>{\textfont0=\the@{#1}{#2}{#3}}c<{\DC@end}}
\makeatother

\usepackage{pgfplots}
\pgfplotsset{width=10cm,compat=1.9}

\usepackage{array}

\newcolumntype{L}[1]{>{\raggedright\let\newline\\\arraybackslash\hspace{0pt}}m{#1}}
\newcolumntype{C}[1]{>{\centering\let\newline\\\arraybackslash\hspace{0pt}}m{#1}}
\newcolumntype{R}[1]{>{\raggedleft\let\newline\\\arraybackslash\hspace{0pt}}m{#1}}

\usepackage{subfiles}

\usepackage{todonotes}

\RestyleAlgo{ruled}


\journal{Energy and Buildings}
\begin{document}
	
\begin{frontmatter}

\title{Recommender systems and reinforcement learning for human-building interaction and context-aware support: A text mining-driven review of scientific literature}

\author{Wenhao Zhang$^{1}$, Matias Quintana$^{2}$, Clayton Miller$^{1, *}$}

\address{$^{1}$Department of the Built Environment, National University of Singapore, 117566, Singapore}
\address{$^{2}$Future Cities Lab Global, Singapore-ETH Centre, 138602, Singapore}

\address{$^*$Corresponding Author: clayton@nus.edus.sg, +65 81602452}

\begin{abstract}
The indoor environment significantly impacts human health and well-being; enhancing health and reducing energy consumption in these settings is a central research focus. With the advancement of Information and Communication Technology (ICT), recommendation systems and reinforcement learning (RL) have emerged as promising approaches to induce behavioral changes to improve the indoor environment and energy efficiency of buildings. This study aims to employ text mining and Natural Language Processing (NLP) techniques to thoroughly examine the connections among these approaches in the context of human-building interaction and occupant context-aware support. The study analyzed 27,595 articles from the ScienceDirect database, revealing extensive use of recommendation systems and RL for space optimization, location recommendations, and personalized control suggestions. Although these systems are broadly applied to specific content, their use in optimizing indoor environments and energy efficiency remains limited. This gap likely arises from the need for interdisciplinary knowledge and extensive sensor data. Traditional recommendation algorithms, including collaborative filtering, content-based and knowledge-based methods, are commonly employed. However, the more complex challenges of optimizing indoor conditions and energy efficiency often depend on sophisticated machine learning (ML) techniques like reinforcement and deep learning. Furthermore, this review underscores the vast potential for expanding recommender systems and RL applications in buildings and indoor environments. Fields ripe for innovation include predictive maintenance, building-related product recommendation, and optimization of environments tailored for specific needs, such as sleep and productivity enhancements based on user feedback. The study also notes the limitations of the method in capturing subtle academic nuances. Future improvements could involve integrating and fine-tuning pre-trained language models to better interpret complex texts.   

\end{abstract}



\end{frontmatter}

\normalsize	

\section{Introduction}

\label{sec:introduction}
\subsection{Background}
The indoor environment significantly affects human health and well-being, with individuals spending on average approximately 86\% of their day indoors \cite{LIU2024114164, Klepeis2001}. The COVID-19 pandemic has further highlighted the critical importance of indoor air quality, propelling research focused on mitigating airborne transmission of pathogens within buildings \cite{AGARWAL2021102942}. For example, \cite{en17112769} has proposed a tailored approach to calculate the optimal number of outdoor air changes in HVAC systems for school buildings to address post-pandemic challenges. However, some studies have also highlighted that improved ventilation systems, while enhancing occupant health, can impact energy consumption and operational costs \cite{ZHENG2021100040, LIN2024970}. Consequently, a critical question in contemporary research is how to balance the dual objectives of improving health and well-being in indoor environments while minimizing energy consumption \cite{Zhang_2022}. In recent years, smart building control has become an important topic in this regard, with extensive studies demonstrating its ability to significantly reduce energy usage while maintaining indoor comfort levels \cite{ZHANG2019472, GAO2023106852}. However, despite these promising results, the adoption rate of smart control technologies remains relatively low \cite{Ejidike2023}. The primary barriers to widespread adoption are the high initial costs and complexity of implementation \cite{10.1007/978-3-030-97748-1_3}. Therefore, there remains a need to explore a more feasible, user-friendly and cost-effective solution. 

In the context of the widespread adoption of Artificial Intelligence (AI), the Internet of Things (IoT) devices, and smart mobile devices, a viable approach is the behavior change facilitated by Information and Communication Technology (ICT) \cite{RSforenergyefficiency, BELHADI202113}. This method leverages recommendation systems and smart devices to influence and change human behavior, thus improving indoor environmental quality and reducing energy consumption \cite{RSforenergyefficiency}. Central to this strategy are sophisticated recommendation algorithms that analyze user preferences and environmental data to provide personalized suggestions. These algorithms have shown success in various sectors, including mobile health, commonly referred to as Just-In-Time Adaptive Interventions (JITAI) in the domain \cite{liao2019personalizedheartstepsreinforcementlearning}, as well as in online shopping \cite{CHOI2012309}, entertainment \cite{10.1145/502585.502625, WANG2014667}, and social networking \cite{Walter2008}, etc. 

In the context of building and human-building interactions, these systems are typically utilized to optimize indoor environments \cite{KIM2024111396, KAR2019135} and enhance building energy efficiency \cite{CUI2016251, s23187974, ALSALEMI2023100741}. Mostly, the methods involve the use of real-time data and federated learning to train models that generate energy saving suggestions \cite{en12071317, KAUR2019288, Varlamis2023}, alongside the integration of sensor data with user preference feedback to provide personalized recommendations \cite{9291639, RIABCHUK2024102559}. Furthermore, reinforcement learning-based recommender systems are widely utilized in this domain due to their ability to handle multi-objective optimization tasks \cite{ONILE2023104392}. For example, multi-agent deep reinforcement learning (RL) has been developed to optimize energy consumption, occupant comfort, and air quality simultaneously in commercial buildings \cite{10.1145/3600100.3623735}. These systems are typically reactive, meaning they respond to user inputs or environmental sensors to provide suggestions. However, proactive recommendation systems have also been explored, which aim to incrementally alter user habits during specific moments based on the theories of \emph{micro-moments} and \emph{habit loops} \cite{8673323}. Furthermore, due to the opacity of machine learning (ML) models \emph{black box}, explainable AI has been used to provide explainable and personalized energy efficiency suggestions \cite{Sardianos_2020}. On an urban scale, the application of recommendation systems has also garnered interest, primarily to improve energy efficiency in smart grids \cite{7997914, GUO2024111219} and facilitate demand response initiatives \cite{BEHL201630}. 

\begin{table*}[h!]
\centering
\renewcommand{\arraystretch}{1.2}
\begin{tabular}{p{0.7cm} p{0.9cm} p{1.4cm} p{4.5cm} p{8.5cm}}
\hline
\textbf{Ref.} & \textbf{Year} & \textbf{Sector} & \textbf{Focus} & \textbf{Contributions} \\ \hline
\cite{HIMEUR20211} & 2021 & Building & 1) algorithm types; 2) algorithm design; 3) applications; 4) limitations; 5) future directions & 1) The first critical review of energy efficiency recommendation systems in buildings; 2) Provides classification and identifies current challenges and unresolved issues, and offers insights into future directions. \\ \hline
\cite{ONILE2021997} & 2021 & Energy services & 1) algorithm types; 2) platforms; 3) user profiles; 4) applications\ & 1) Reviews recommendation system's applications in demand-side management, smart services, e-commerce, residential, industrial, commercial, and policy contexts. \\ \hline
\cite{Mohammadi2019} & 2019 & IoT & 1) literature quantification analysis; 2) system design; 3) limitations; 4) future directions & 1) Examines recommendation techniques in IoT environments; 2) Highlights advantages, disadvantages, and unresolved challenges for various methods. \\ \hline
\cite{10.1093/comjnl/bxab049} & 2021 & IoT & 1) literature quantification analysis; 2) algorithm types 3) applications; 4) algorithm design; 5) future directions & 1) Comprehensively reviews IoT-based recommendation technologies; 2) Summarizes applications in various fields and identifies challenges; 3) Provides an RSIoT reference framework to guide future research and practices. \\ \hline
\cite{QUIJANOSANCHEZ2020101545} & 2020 & Smart cities & 1) algorithm types; 2) system design; 3) applications; 4) literature quantification analysis & 1) Reviews the status and key drivers of recommendation systems for smart cities across domains like economy, environment, mobility, and living. \\ \hline
\cite{smartcityreview1} & 2023 & Smart cities & 1) literature quantification analysis; 2) applications; 3) future directions & 1) Highlights recommendation systems in energy efficiency, government services, healthy living, parking optimization, and traffic congestion reduction. \\ \hline
\cite{electronics13112151} & 2024 & Smart cities & 1) literature quantification analysis; 2) algorithm types; 3) applications & 1) Explores trends, themes, and collaborations in smart city recommendation systems; 2) Identifies cooperation between authors and institutions. \\ \hline
\cite{electronics13071249} & 2024 & Smart cities & 1) literature quantification analysis; 2) algorithm types; 3) applications & 1) Summarizes recommendation systems in tourism, health, mobility, and transport; 2) Highlights common algorithms and application scenarios. \\ \hline

\end{tabular}
\caption{Summary of Literature on Recommendation Systems}
\label{table:recommendation-systems}
\end{table*}

\subsection{Previous reviews and research gaps}

Recommendation systems have been widely applied to improve various aspects of building performance and energy management and several literature reviews have recently been conducted to summarize the applications of recommendation systems in the context of building and the built environment. These reviews analyze current studies on recommendation systems in various areas, including energy efficiency \cite{HIMEUR20211, ONILE2021997}, Internet of Things (IoT) \cite{Mohammadi2019, 10.1093/comjnl/bxab049}, and smart cities \cite{QUIJANOSANCHEZ2020101545, smartcityreview1, electronics13112151, electronics13071249}. Table \ref{table:recommendation-systems} summarizes the focus areas and key contributions of these review articles. From this table, it is evident that most reviews analyze the application of recommendation systems in the field of building and the built environment within the context of smart cities and the IoT, and only one study has reviewed their application at the building scale. However, this study exclusively focuses on the use of recommendation systems for building energy efficiency \cite{RSforenergyefficiency}. It does not explore their potential contributions to other core objectives of building-human interaction design, such as improving occupant wellness and comfort. Furthermore, these reviews focus primarily on analyzing the individual components of recommendation systems (typically data sources, algorithm types, system types, platforms, and recommendation objectives) in isolation. They fail to investigate the interrelationships among these categories, such as the connections between different objectives and data types or algorithms, which is crucial for guiding the design of recommendation systems tailored to specific objectives. This issue often arises due to the limitations of human analytical capacity, as it is challenging to discern subtle relationships between different studies, particularly in interdisciplinary fields \cite{KIM2022101255, Mah}. Therefore, to address this gap and investigate the interrelations between various components, a more sophisticated approach is essential to efficiently manage and interpret extensive literature datasets.

\subsection{Text mining-based literature review}
Text mining-based literature reviews utilize data-driven technologies to systematically extract and analyze information from a large corpus of text, which is particularly advantageous in fields characterized by rapid technological advancements and extensive academic output \cite{O’Mara-Eves2015}. Using natural language processing (NLP) and ML algorithms, text mining can efficiently process text data, including classification, clustering, and association, to identify trends, patterns, and emerging themes that might not be evident through traditional review methods \cite{LimandMaglio}.

Previous applications of text mining in literature reviews have demonstrated promising results. Tools such as VOSviewer \cite{vaneck2011textminingvisualizationusing} are commonly used in bibliometric analysis and network visualization \cite{PARK20182664}, using paper metadata to create bibliometric networks and density maps between articles in various fields \cite{YAN2022e12088, WUNI201969}. However, these tools are often constrained by graphical user interfaces (GUIs), which limit the user's ability to modify or extend beyond the provided algorithms \cite{Mah}. To address specific text mining needs, researchers have adopted advanced ML techniques such as topic modeling. For example, Latent Dirichlet Allocation (LDA) \cite{10.5555/944919.944937} and the Correlated Topic Model (CTM) \cite{10.5555/2976248.2976267} have been effectively used to model documents as mixtures of topics \cite{Chen2015WhatOC, RePEc:kap:jbuset:v:139:y:2016:i:1:d:10.1007_s10551-015-2622-4, MH90246}. Similarly, \cite{10323487} employs the BERTopic model \cite{grootendorst2022bertopicneuraltopicmodeling} to efficiently extract and cluster topics, thus providing deeper insights into textual data. Recent advances in computational power and algorithmic sophistication have also led to the emergence of Large Language Models (LLMs), such as GPT \cite{brown2020languagemodelsfewshotlearners} and LLaMA \cite{touvron2023llamaopenefficientfoundation}, offering an alternative to traditional topic modeling approaches \cite{mu2024largelanguagemodelsoffer}. Furthermore, word embedding technologies such as Word2Vec \cite{NIPS2013_9aa42b31} and GloVe \cite{pennington-etal-2014-glove} facilitate character-level analysis \cite{Mah}. These are complemented by tools such as Python and Scikit-Learn \cite{10.5555/1953048.2078195}, RapidMiner Studio \cite{10.1145/1150402.1150531}, N-gram \cite{10.5555/176313.176316}, and the Natural Language Toolkit (NLTK) \cite{loper2002nltknaturallanguagetoolkit} for pre or post-processing, offering a nuanced understanding of word relationships and topic significance \cite{Bauer2021, Gurcan2020ResearchTO}.

Among these methods, Word2Vec is frequently used by researchers to quantify the relationships between terms associated with different topics from large volumes of literature. For instance, \cite{Tshitoyan2019} developed an unsupervised embedding model based on Word2Vec using abstracts from 3.3 million papers to extract materials science information and predict new thermoelectric materials; \cite{Bauer2021-mn} utilized abstracts from nearly 4 million publications as training data to explore the relationships between cancer types and anticancer drugs; and  \cite{ZHU2023106876} employed 4,633 articles retrieved from the Web of Science database to develop a Word2Vec model, uncovering latent patterns and interactions within resources, conservation and recycling research topics. In the context of building energy efficiency, \cite{Mah} retrieved approximately 30,000 scientific publications via the Elsevier API to extract relationships among data sources, data science techniques, and applications for building energy efficiency across the entire lifecycle of buildings. This study represents the first attempt at such a text mining and NLP-based review in the field of building energy efficiency research.

\subsection{Review of recommender systems for human-building interaction using text mining}
\begin{figure*}[h] 
    \centering
    \includegraphics[width=\textwidth]{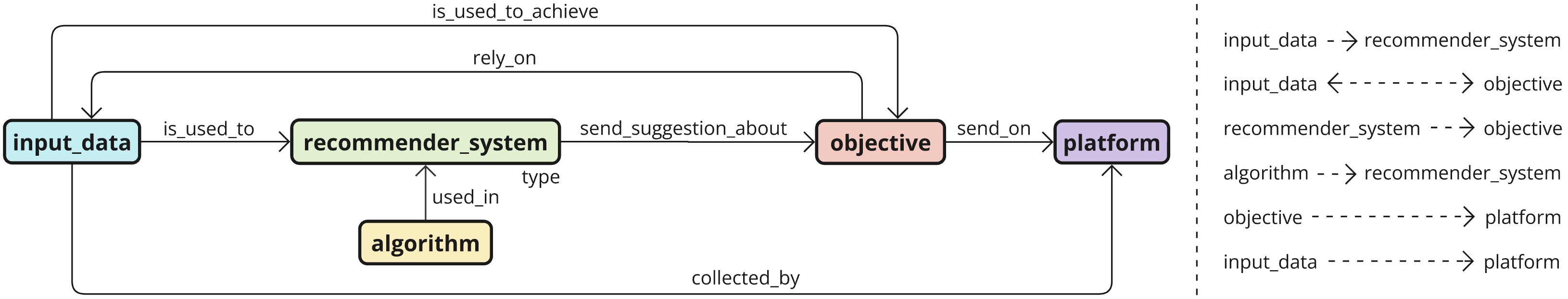}
    \caption{Overview of the categories of concepts analyzed in this text mining analysis and their relationships with each other.}
    \label{fig:objective}
\end{figure*}

This study aims to leverage data mining techniques proposed by Abdelrahman et al. \cite{Mah} to comprehensively review the applications of recommendation systems in the context of human-building interaction and occupant context-aware support. Specifically, it seeks to quantify the interrelationships among five key components of recommendation systems: \emph{algorithms}, \emph{types of recommendation systems}, \emph{input data}, \emph{interventions/objectives}, and \emph{platforms}. Figure \ref{fig:objective} illustrates six specific relationships among these five categories, which have been identified through prior literature reviews. 

The objectives of this study are as follows: 1) to train a Word2Vec model on relevant academic papers to extract keyword relationships; 2) to create heatmaps that visualize the interrelationships between keywords within the different components of recommendation systems; 3) to analyze the strengths and weaknesses of these relationships as shown in the heatmaps, thereby identifying the current state of development in the field. The main contributions of this study can be summarized as follows:
\begin{enumerate}
    \item This study is the first to provide a comprehensive review of recommendation systems in the context of human-building interaction and occupant context-aware support.
    \item Quantifies the correlations between recommendation objectives, input data, algorithms, and platforms by using NLP and text mining techniques.
    \item The findings offer practical guidance on selecting appropriate data and algorithms to design recommendation systems that target different objectives.
    \item Explores the most widely used recommendation algorithms, application areas, and future development directions.
\end{enumerate}

\section{Methodology}
\label{sec:methodology}

\begin{figure*}[h] 
    \centering
    \includegraphics[width=\textwidth]{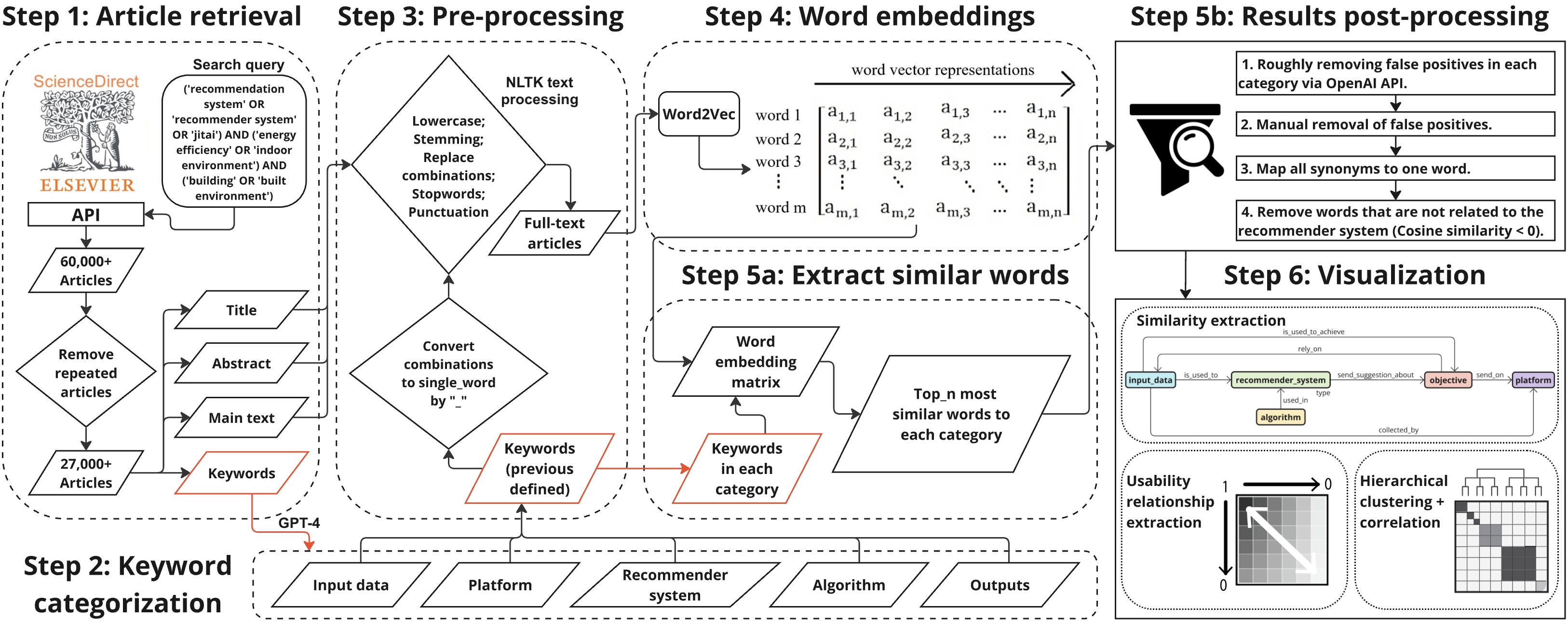}
    \caption{The flowchart shows the methodology used in this research: 1) Identifying the querying keywords of each category, 2) Extracting the relevant articles with their corresponding metadata using Elsevier API, 3) Pre-processing the data, 4) Applying the NLP algorithms, 5) Extracting the relationships among these categories, 6) Similarity matrix visualization.}
    \label{fig:methodology}
\end{figure*}

The methodology adopted in this study integrates several phases, namely text mining, NLP and visual analytics, which are structured into a systematic six-step process, as illustrated in Figure \ref{fig:methodology}. This comprehensive approach encompasses the following stages: 1) collecting relevant papers via the Elsevier API, 2) extracting and identifying keywords of each category, 3) preprocessing the articles for NLP model training, 4) applying NLP techniques to generate numerical text representations, 5) extracting the relationships among these categories, 6) Creating the relation graph network. 

\subsection{Step 1: Article retrieval}
ScienceDirect, operated by Elsevier, is a comprehensive academic database and full-text library that serves as the main source for scholarly articles \cite{TOBER2011139}. This study uses the Elsevier API \cite{VanNoorden2014} to perform full-text searches within the ScienceDirect database. The full text of each article and key metadata, including titles, abstracts, keywords, authors, and publication dates, are extracted and stored in JSON format. This format supports efficient data manipulation and retrieval for subsequent analysis phases.

\subsubsection{Search query}
The search strategy is designed to extract articles related to the application of recommender systems in buildings and occupant interactions. The query utilized combines specific terms to refine the focus of the retrieval: (\emph{recommendation system} OR \emph{recommender system} OR \emph{jitai}) AND (\emph{energy efficiency} OR \emph{indoor environment}) AND (\emph{building} OR \emph{built environment}). The inclusion of terms (\emph{building} OR \emph{built environment}) ensures that the search results are relevant to the fields of building and built environment studies.

\subsubsection{Article filtering}\label{sec:article-filtering}
\begin{figure*}
    \centering
    \includegraphics[width=1\textwidth]{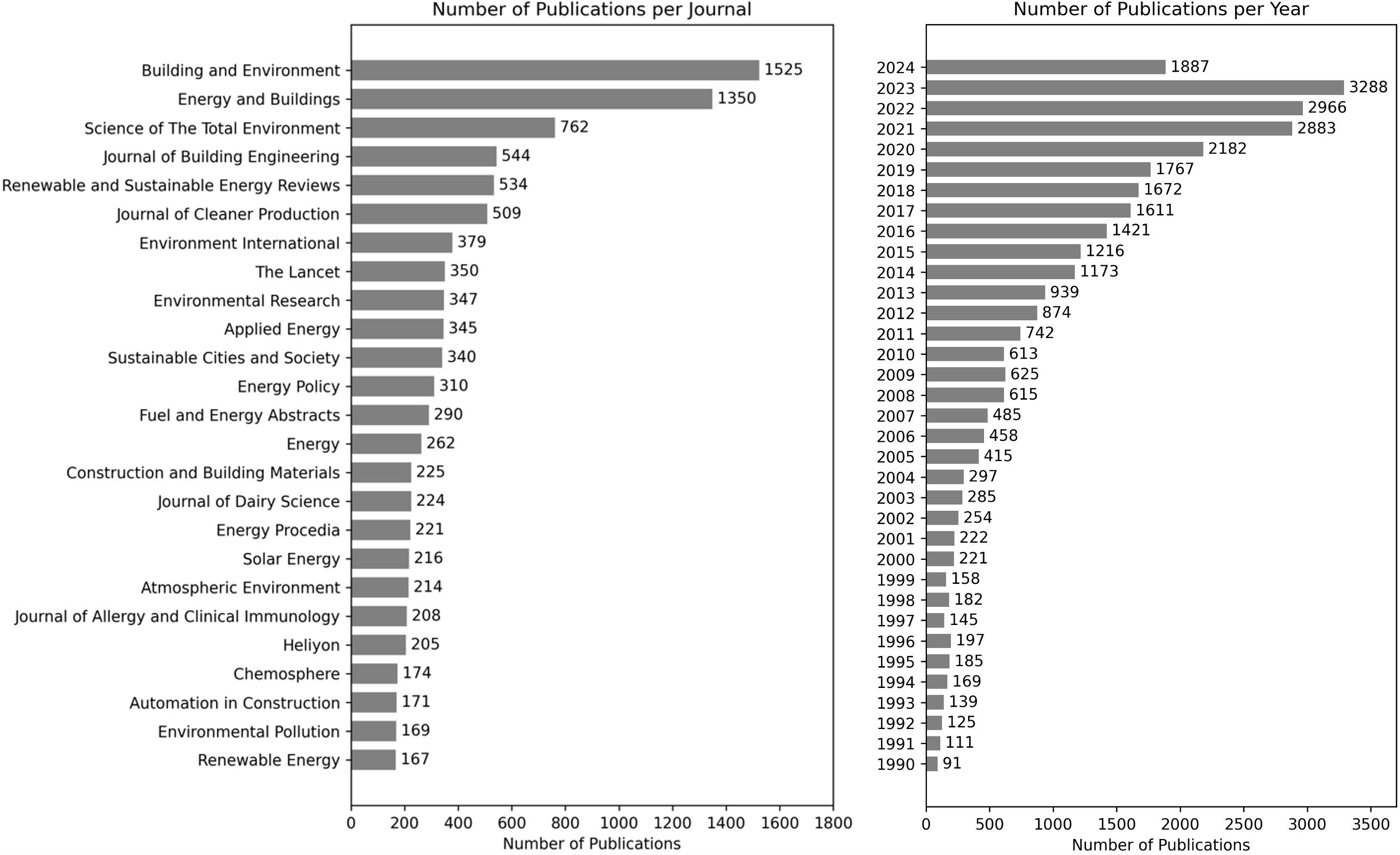}
    \caption{Number of collected papers per journal and per year.}
    \label{fig:publication}
\end{figure*}

The initial query procedure yielded more than 60,000 articles sourced from more than 1,000 journals (the initial inquiry was conducted in May 2024). A significant proportion of these articles were found to be duplicated and some contained incomplete or erroneous information. Following the removal of these duplicates and erroneous files, the data set was refined to 27,595 articles. Each retained article includes the full text along with essential metadata such as the author(s), publication year, title, abstract, and journal of publication. Figure \ref{fig:publication} displays the distribution of the number of articles per journal and the trends of annual publication. Analysis of these data reveals that most of the articles originate from journals focused on the construction, energy, and environment sectors. In addition, a smaller subset of these articles is derived from medical journals. 

\subsection{Step 2: Keyword categorization}

Upon successful retrieval and download of articles, keywords from each article are extracted and categorized into the five predefined categories outlined in Section \ref{sec:definitions}, which are essential for subsequent similar word extraction processes. The compound keywords are then transformed into single entities to facilitate the training of NLP models, with constituent words linked by ``\textunderscore''. This modification aims at preserving semantic integrity and will be detailed in the text processing section. For the Named Entity Recognition (NER) task, LLMs have demonstrated commendable accuracy in performing such tasks, as evidenced by existing research \cite{10.1093/jamia/ocad259, 10497008}. Specifically, this study uses the OpenAI GPT-4 model \cite{openai2024gpt4technicalreport} to carry out NER. A manual review will be conducted following the initial automated classification performed by the LLM. 

\subsubsection{Definitions}\label{sec:definitions}
Detailed definitions for each category mentioned in the objectives section are provided below: 

\begin{enumerate}[label=Def.\arabic*:, leftmargin=*]
    \item \textbf{Input data:} (\emph{input\textunderscore data}) refers to various types of input data employed in recommendation systems. This includes building and environmental data, such as indoor temperature, humidity, and indoor environmental quality; physiological data, such as heart rate, body temperature, and activity patterns; and user context data, such as user profiles and real-time feedback.
    \item \textbf{Recommender system:} (\emph{recommende\textunderscore system}) refers to the types of recommender systems commonly used in the field of human-building interaction. This encompasses systems like JITAI, context-aware recommender systems for dynamic user environments, and RL-based recommender systems that adapt based on user interactions.
    \item \textbf{Algorithm:} (\emph{algorithm}) refers to the types of algorithms employed across various recommender systems. Notable examples include NLP, deep learning, and RL.
    \item \textbf{Objective:} (\emph{objective}) refers to the goals pursued by recommender systems under the context of human-building interaction. Examples include promoting energy-saving behavior, altering sedentary behaviors, or providing personalized control recommendations.
    \item \textbf{Platform:} (\emph{platform}) refers to both the platforms used to collect input data for recommender systems and the platforms on which system outputs are delivered. This encompasses smart wearable devices, as well as traditional computing platforms such as computers and smartphones.
\end{enumerate}

Each of these categories was manually selected from a preliminary survey of the existing literature and extraction of pertinent terms from these texts. At this stage, the extracted articles were ready for preprocessing for the subsequent model training stage.

\subsection{Step 3: Text pre-possessing}
The pre-processing stage is designed to refine the full-text data for subsequent data mining processes. The preparation involves removing words that could negatively impact model training, merging compound words, and performing word tokenization and lemmatization. Initially, two types of undesirable words are identified: 1) metadata elements such as article or image IDs, author information, hyperlinks, and annotations, including terms such as \emph{introduction}, \emph{methodology}, \emph{reference}, \emph{figure}, and \emph{table}, which are repetitive between articles and can introduce bias in subsequent analyses; 2) stop words, which occur frequently but contextually insignificant words such as \emph{the}, \emph{a}, \emph{in}, and \emph{of}, which, if included, could bias the NLP models. NLTK is then used to remove these stop words \cite{loper2002nltknaturallanguagetoolkit}. 

Subsequently, the established dictionary of compound words is used to replace compound terms in the articles. This approach allows the model to be trained solely on 1-grams, eliminating the need for additional training of n-gram models to learn these compound terms. An n-gram model refers to a sequence formed by contiguous n words, capable of capturing the relationships between these consecutive words \cite{articlengram}. Typically, training an n-gram model requires preprocessing the text into groups of n words, and for different values of n, the model must be re-trained. For example, to train the NLP model to understand the term \emph{building energy model}, which constitutes a 3-gram, the input text must be tokenized into groups of three words each for model training. Therefore, after the compound word merging, all n-grams are converted into 1-grams (a single word). This means that each compound word is treated as an independent unit, which simplifies the processing workflow and significantly reduces the time required for model training. Through this method, there is no longer a need to train a dedicated n-gram model to recognize combinations of words. Although this approach only includes words appearing in the keyword section of the papers, it is sufficient to cover the major abbreviations and definitions in the literature, since the keyword section typically contains the core concepts and terms of the articles.

In the final stage, tokenization is performed at the sentence level for each word within the sentences. Since NLP models are case-sensitive, it is necessary to perform lemmatization and stemming to ensure word uniformity. For example, in NLP models, the same word in uppercase and lowercase is treated as distinct entities. To address this, all words are first converted to lowercase. Subsequently, text lemmatization will be employed to extract the common roots of different word forms. After the full text is standardized, it is prepared for the NLP text mining process.

\subsection{Step 4: NLP text mining using Word2Vec}
Word2Vec is a Python-based word embedding tool that is used to identify semantic similarities between words within a text \cite{mikolov2013efficientestimationwordrepresentations}. During the Word2Vec model training, words are projected into a vector space based on their semantic context within surrounding text. Words with similar meanings tend to be positioned closer to each other in this vector space. The relationships between these words are then quantified using the cosine similarity between their respective vectors (Equation \ref{eq:cosine_similarity}).

\begin{equation}
\text{sim}(A, B) = \cos(\theta) = \frac{A \cdot B}{\|A\| \|B\|}
\label{eq:cosine_similarity}
\end{equation}

The objective of training with Word2Vec is to identify an optimal vector representation for each word in the vector space that reflects its semantic meaning. Specifically, Word2Vec utilizes context within sentences, defined as a window that encompasses n words surrounding the target word, to predict the target word or the center word. During training, each word in a sentence is sequentially designated as the center word. The vector representations for both the center word and its surrounding context words are updated based on the discrepancy between the predicted center word vector and its actual vector. Training continues until these discrepancies no longer significantly decrease or model convergence. Convergence implies that the vectors calculated based on surrounding words are sufficiently close to the actual vector of the center word or that further training does not significantly enhance model performance. 

In this study, the preprocessed full texts of approximately 27,595 articles were used as training data for the Word2Vec model. Before training the model, hyperparameter configuration was conducted. Specifically, the \emph{min\_count} parameter was set to 2 to remove words with a frequency of less than 2, thereby reducing noise. To optimize the remaining hyperparameters, including \emph{vector\_size} and \emph{window}, a synonym list containing 50 pairs of domain-specific terminology was created. This synonym list was developed based on a prior literature review. Hyperparameter tuning was conducted by evaluating the similarity between each pair of terms in the list under different hyperparameter combinations. The objective function was the average similarity of terms in the list(Equation \ref{eq:objective_function}), a higher average similarity indicated better model performance.

\begin{equation}
\text{Objective function} = \frac{1}{50} \sum_{i=1}^{50} sim(A_i, B_i)
\label{eq:objective_function}
\end{equation}

After testing 25 hyperparameter combinations, the optimal model was identified with a context window of 20 words and a vector size of 300, achieving a maximum overall similarity of 0.714. Furthermore, the study conducted analogy tests to evaluate whether the model successfully learned the semantic relationships between words. The principle behind these tests involves performing vector operations on words with specific semantic relationships to derive other words with analogous relationships. For example, in the model, the operation \emph{hvac $-$ thermostat $+$ lighting} results in the vector corresponding to \emph{dimmer}. This indicates that the model effectively captured the semantic relationship between the input words, as the relationship between \emph{hvac} and \emph{thermostat} parallels that between \emph{lighting} and \emph{dimmer}.

\subsection{Step 5: Extracting the relationship between categories}
After the training process, the trained Word2Vec model can be used to identify words related to the input word based on their proximity in the vector space, and compute the similarity between two input words by calculating the cosine similarity between their respective word vectors. Before extracting the correlation between two words, predefined words within each category are used as input to the NLP model to extract words that are semantically similar. This procedure aims to identify and retrieve all relevant words associated with each category from the existing literature. For each word, the model extracts the top 100 similar words. Subsequently, any terms deemed irrelevant are removed using the GPT-4 model. This step ensures that the final dataset is refined and relevant to the specific categories under study. After this preliminary filtering, any remaining words are subjected to manual review and further refinement by human analysts. 

Following the extraction of keywords, it is often observed that multiple words may refer to the same concept. In such instances, these words are mapped to a single representative term. For example, the term \emph{smart\_phone} is associated with several synonyms, including \emph{android\_smartphone}, \emph{iphone}, \emph{smartphone}, and \emph{smart-phone}. Variations among these synonyms can involve minor differences, such as the use of underscores (\emph{"\textunderscore"}) instead of hyphens (\emph{"-"}) or the substitution of \emph{stage} for \emph{phase}. Another example is the use of acronyms that refer to the same term, such as \emph{jitai} and \emph{just-in-time\_adaptive\_intervention}. Such synonyms and their relationships can be efficiently identified through the similarity matrix in Word2Vec. After compiling a comprehensive list of words and their synonyms from identified categories, the relationships among them are analyzed. The final similarity measure for each list in relation to other words is determined by calculating the average similarity of each word in the list to a target term. This approach ensures a systematic and accurate representation of the terminological relationships within the data set.

\section{Results}
\label{sec:results}
Table \ref{tab:recommendation_systems} displays the types of recommender systems extracted from the literature and their description. For words extracted in each category, those with a relevance score of less than zero to these recommender system types were eliminated, ensuring that all retained terms positively correlate with the recommender systems.

\FloatBarrier
\begin{table*}[h]
\centering
\begin{spacing}{1.2} 
\begin{tabular}{>{\raggedright\arraybackslash}m{5cm}>{\raggedright\arraybackslash}m{12cm}}
\hline
\textbf{Type} & \textbf{Description} \\
\hline
Content-based system & A system that recommends products similar to the user's past favorites based on feature comparisons \cite{Lops2011}. \\
\hline
Collaborative filtering-based system & A system that utilizes the ratings or behavior of a group of users to make recommendations to similar users \cite{Schafer2007}. \\
\hline
Knowledge-based system & A system that uses domain-specific knowledge to provide recommendations \cite{Knowledge-Based}. \\
\hline
Hybrid recommendation system & Combines two or more types of recommender systems to reduce the limitations of any single approach \cite{Burke2002}. \\
\hline
Social recommendation system & Leverages users' social connections and interactions to recommend items \cite{Walter2008}. \\
\hline
Context-aware recommendation system & Incorporates contexts such as time, location, or environment sensor data to offer more appropriate recommendations \cite{1423975}. \\
\hline
Reinforcement learning-based system & A system that dynamically adapts to user preferences over time by continuously learning from user interactions to maximize a reward function  \cite{712192}. \\
\hline
Just-In-Time Adaptive Interventions (JITAI) &  A system designed to provide intervention on an as-needed basis, adapting in real-time to an individual's changing circumstances and conditions \cite{PMID:27663578}. \\
\hline
\end{tabular}
\caption{Definition of different types of recommendation systems}
\label{tab:recommendation_systems}
\end{spacing}
\end{table*}

\begin{figure*}[h]
    \centering
    \includegraphics[width=1\textwidth]{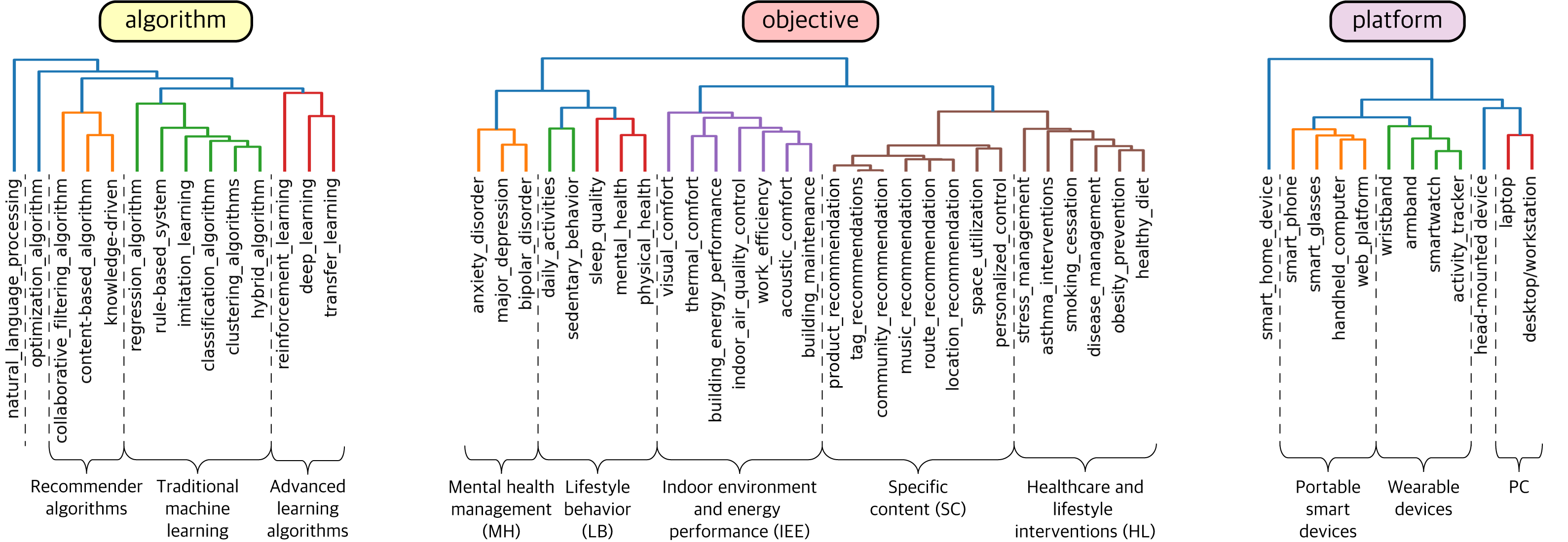}
    \caption{The hierarchical agglomerative clustering (HAC) of the objective, algorithm, and platform.}
    \label{fig:out_alg_plt_cluster}
\end{figure*}
    
\begin{figure*}[h]
    \centering
    \includegraphics[width=1\textwidth]{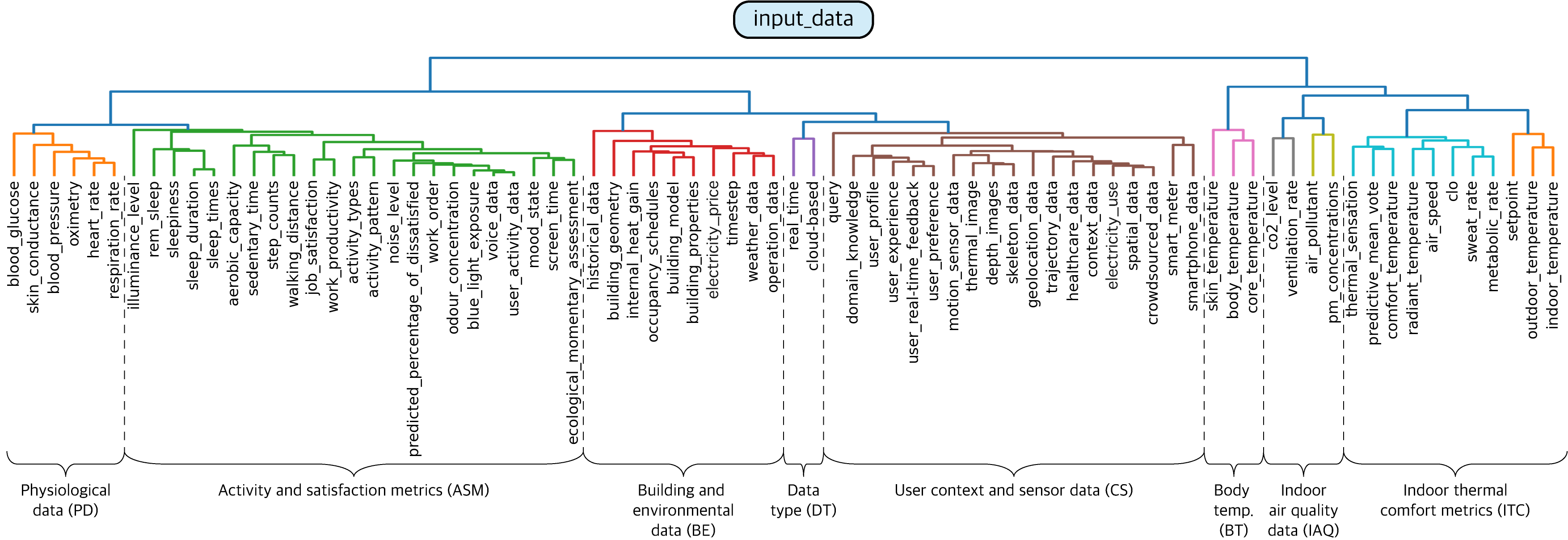}
    \caption{The hierarchical agglomerative clustering (HAC) of the input data.}
    \label{fig:in_cluster}
\end{figure*}

In this analysis, heat maps are utilized to demonstrate the relevance between pairs of words. The intensity of the heat map's color represents the strength of the relationship: darker colors indicate stronger relationships, while lighter colors suggest weaker ones. These heat maps not only provide a quantitative representation of correlation but also serve as a tool to guide design decisions. For example, in heat map showing the relationship between input data and design objectives, the strength of the connections can help identify the necessary input data to achieve specific objectives, as well as irrelevant data types. Similarly, heat map depicting the relationship between recommendation system types and objectives can help determine the most suitable system type for a given design objective. Additionally, hierarchical clustering is used to form sub-clusters within each category, which serve to present the compositional structure of each category and enhance the analysis of interrelationships among these clusters. Specifically, Hierarchical Agglomerative Clustering (HAC) utilizing Ward's method was applied to the word embedding vectors within each category. This approach clusters similar words based on their Euclidean distances in the vector space, facilitating the representation of these clusters in a tree structure known as a dendrogram. Figures \ref{fig:out_alg_plt_cluster} and \ref{fig:in_cluster} illustrate the data mining results and subclusters within the categories of input data, objectives, algorithms, and platforms.

\subsection{Input data for different recommendation objectives}

The initial comparison provides an overview of the relationship between input data and output in recommender systems, focusing solely on understanding the types of input data required for different outputs. In this analysis, both axes are organized based on the average strength of relationships, with the horizontal axis representing the objectives of the recommender systems (arranged from left to right) and the vertical axis detailing the clusters of input data (arranged from top to bottom).

It can be seen that data are predominantly utilized to enhance indoor comfort and energy performance (IEE), as well as to provide specific content (SC) recommendations. However, data are underused in areas such as health care and lifestyle interventions (HL). On the one hand, some recommender systems frequently utilize data such as temperature settings, work efficiency, daily activities, sedentary behavior, and sleep quality. However, applications that rarely use data include those focused on product recommendations and lifestyle and healthcare interventions.

\begin{figure*}
    \centering
    \includegraphics[width=0.7\textwidth]{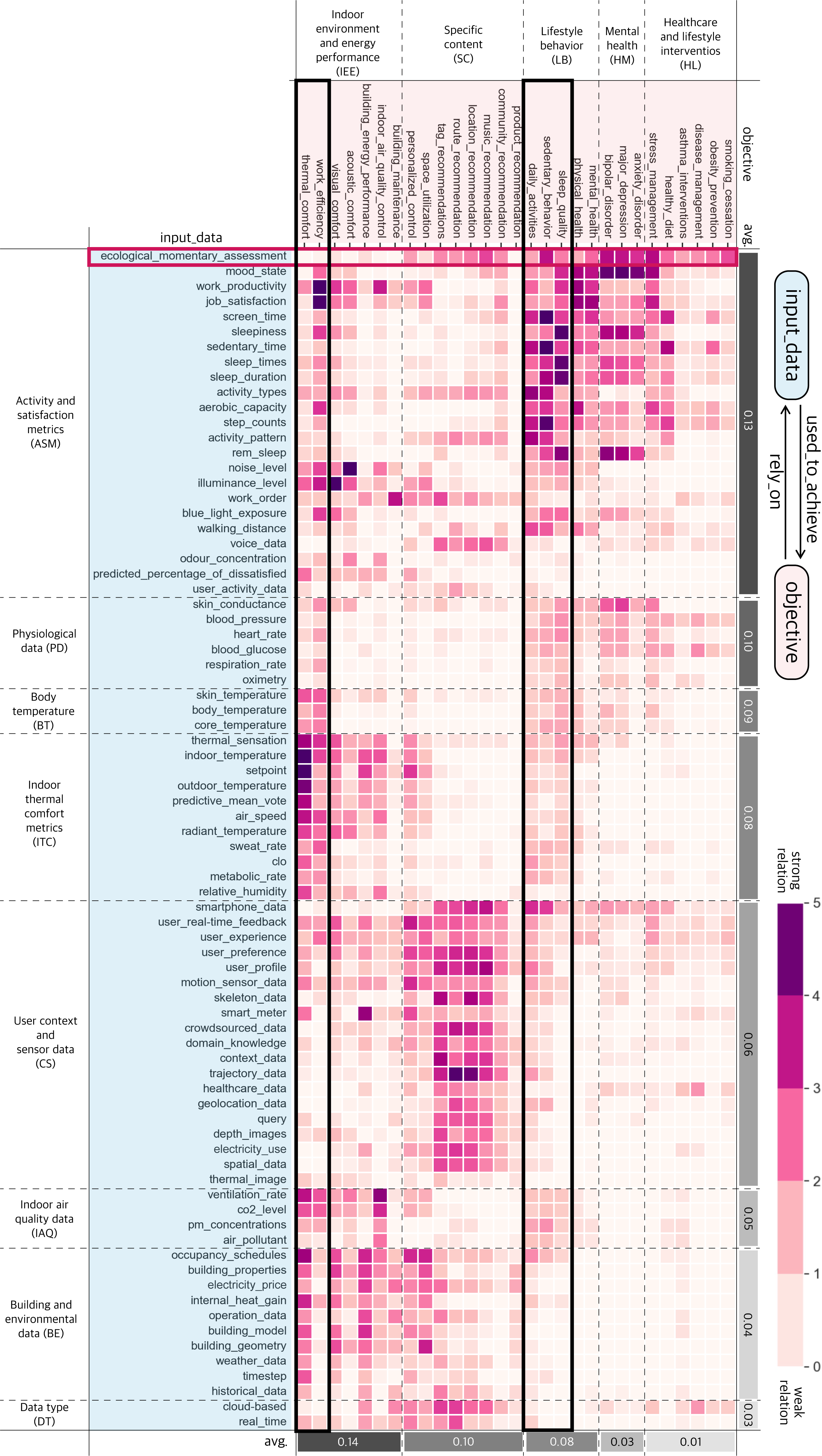}
    \caption{The relationships between input data (shown on the Y-axis with blue highlight) and objectives (shown on the X-axis with red highlight). Both are grouped based on the HAC and sorted based on the sum of similarities per cluster and per row/column in their clusters.}
    \label{fig:first_paper_result_2}
\end{figure*}

Figure \ref{fig:first_paper_result_2} also reveals diversity in the use of data sources. A key finding is the significant role of ecological momentary assessment data, which is shown to be highly relevant across multiple categories of recommender system objectives. This type of data is crucial as it captures real-time feedback from users, enabling recommendation systems to provide timely and context-specific suggestions that optimize user behavior and system efficiency. Other frequently used data types include smartphone data, mood state, skin temperature, and work productivity. Nonetheless, certain data remain underutilized, including physiological measures (PD) such as heart rate and respiration rate; indoor thermal comfort metrics (ITC) such as clo factor, predicted percentage of dissatisfaction, and metabolic rate; as well as environmental sensor data like air pollutants, weather data, thermal image, and relative humidity.

\FloatBarrier

\begin{figure*}[ht]
    \centering
    \includegraphics[width=0.9\textwidth]{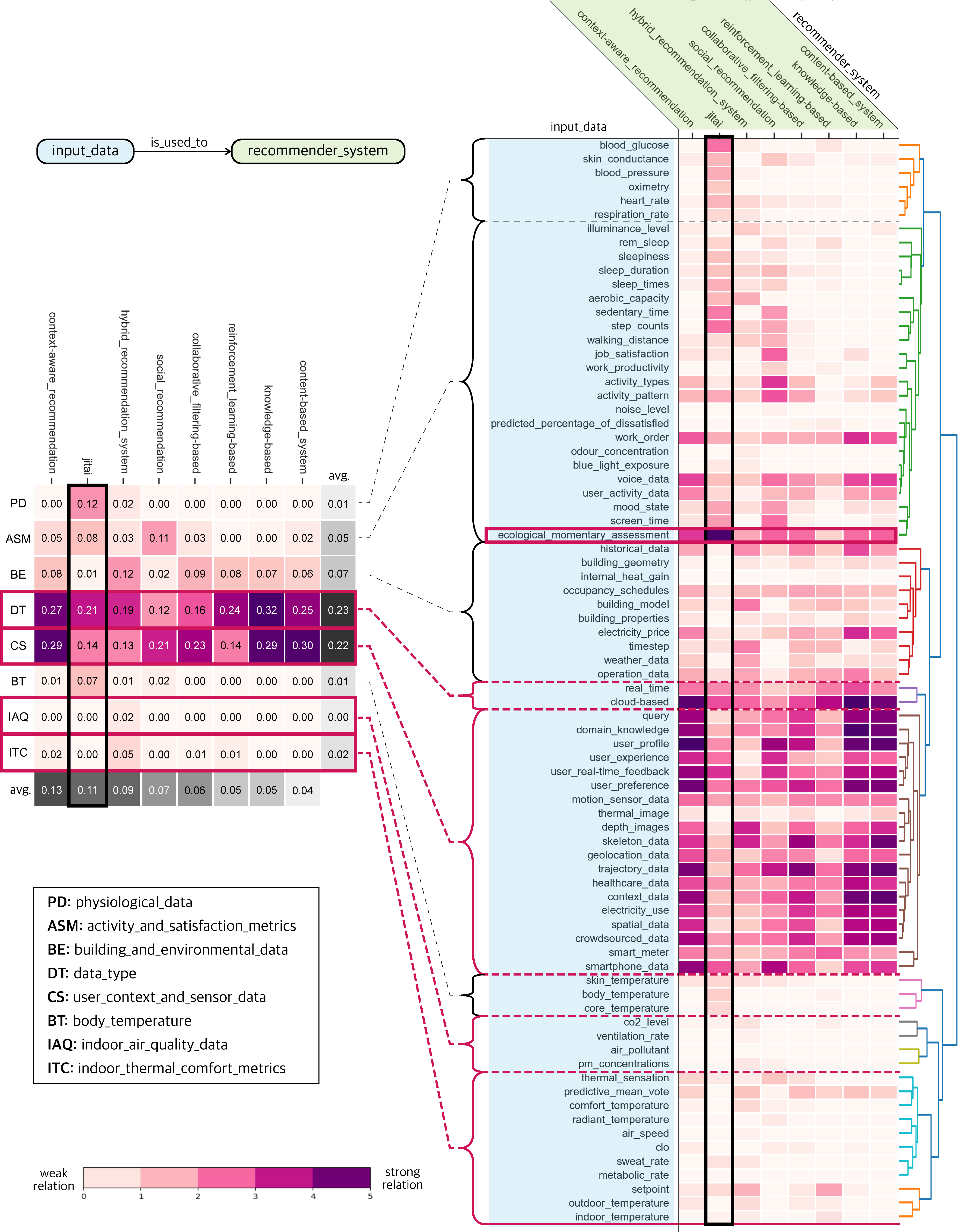}
    \caption{The relationships between input data and different types of recommender systems. The HAC (right) between recommender system type (shown on the X-axis with green highlight) and input data (shown on the Y-axis with blue highlight). The heatmap (left) is a summarized version of the relation map by taking the average of each input data cluster.}
    \label{fig:first_paper_result_1_in}
\end{figure*}

\FloatBarrier
\newpage

\subsection{Input data and objectives of recommender systems for human-building interaction and occupant support}

The subsequent category extracted the strength of relationships between various input data/objectives and different types of recommender systems. The horizontal axis (Recommender Systems) categorizes the types of recommender systems commonly used in human-building interaction and occupant support, as identified in the literature. The vertical axis (Input Data/Output Objectives) is organized by categories of terms. The left diagram presents the average relevance of each category of output terms to the recommender systems, while the right diagram provides a detailed classification of outputs.

Figure \ref{fig:first_paper_result_1_in} highlights the significant relevance of real-time and cloud-based data (DT) in this domain, which underscores the sector's high demand for real-time data, with the majority of recommender system algorithms computed on cloud platforms \cite{RSforenergyefficiency}. The most pertinent data categories for these recommender systems are user context and sensor data (CS). These data types are crucial, as they directly enhance the personalization of recommendations, leveraging real-time environmental and user-specific information to optimize system output. JITAI particularly benefits from high correlations with physiological data (PD) and ecological momentary assessment to provide interventions related to indoor health and behavioral modifications. However, data related to indoor air quality and comfort (IAQ \& ITC), such as indoor pollutant concentrations, carbon dioxide levels, ventilation rates, indoor temperatures, noise levels, predicted percentage of dissatisfied occupants, and clothing insulation factor, exhibit lower correlations with recommender systems. These data types, despite being crucial for the indoor environment quality, are not sufficiently utilized in current recommender systems.

\begin{figure*}[ht]
    \centering
    \includegraphics[width=0.95\textwidth]{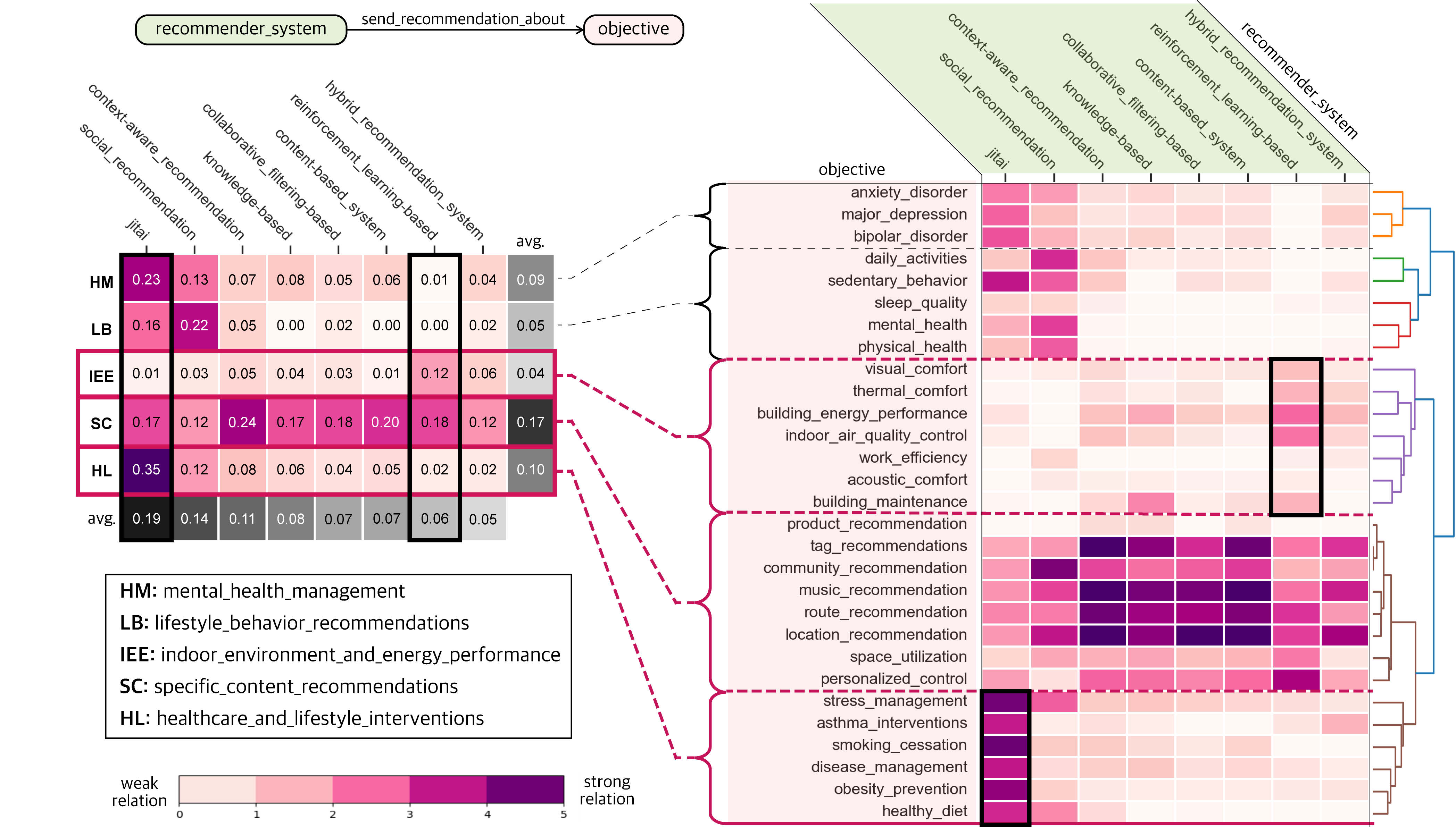}
    \caption{The relationships between objectives and different types of recommender systems. The HAC (right) between recommender system type (shown on the X-axis with green highlight) and objectives (shown on the Y-axis with red highlight). The heatmap (left) is a summarized version of the relation map by taking the average of each objective cluster.}
    \label{fig:first_paper_result_1_out}
\end{figure*}

Figure \ref{fig:first_paper_result_1_out} provides significant insight into how different types of recommender system align with objectives pertinent to the control of the building and the interaction of the occupants. Regarding the types of recommender systems, it highlights that JITAI is the most widely applicable form of recommendation system within this domain, particularly effective in delivering healthcare and behavioral change interventions (HL). However, RL-based recommender systems are not as widely used (showing the lowest relevance to the objectives), but are more closely associated with interventions in indoor environmental comfort and energy performance (IEE). This limited application may be due to the complexity involved in these areas. Regarding the objectives of recommender systems, specific content recommendations (SC), which include suggestions on music, routes, and locations, demonstrate strong associations across all types of recommender systems. However, the analysis reveals a significant gap in the application of recommender systems to objectives directly related to indoor environmental comfort and energy efficiency (IEE). Although these are primary objectives of current building design, recommender systems are currently underutilized in these areas.
\FloatBarrier

\subsection{Data collection and delivery platform used in recommender systems}

\begin{figure*}
    \centering
    \includegraphics[width=0.7\textwidth]{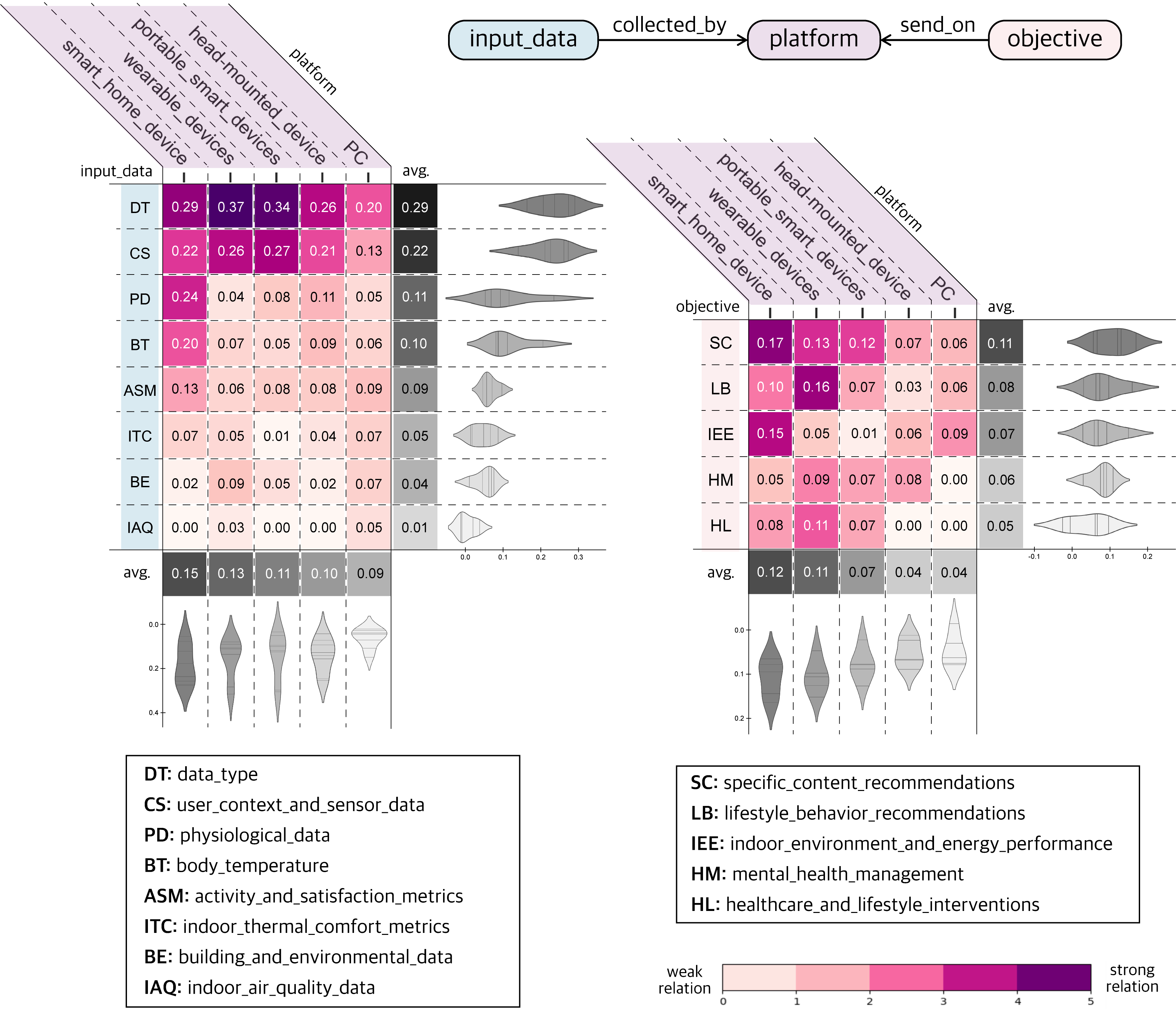}
    \caption{The heatmap on the left shows the average correlation between each pair of elements from the input data and different platforms. The heatmap on the right shows the average correlation between each pair of elements from the objectives and different platforms.}
    \label{fig:first_paper_result_4}
\end{figure*}

The final visualization in this section focuses on the platforms used in the recommender system. The heat map in Figure \ref{fig:first_paper_result_4} illustrates the relationships between the input data types / objectives and the platforms of the recommendation systems. The left side of the figure shows the correlation between the input data and the platforms used to collect these data. Wearable devices and smart home devices exhibit strong correlations with most data types. Head-mounted devices (HMDs), as a new technology, also show a strong correlation with the data. PCs, although capable of collecting building operational data, exhibit weaker connections, possibly due to their limited interaction capabilities and lack of portability.

On the right side, the heatmap illustrates the platforms commonly used to deliver specific content (SC) or behavior interventions (LB). In this domain, smart home devices and wearables remain the primary platforms for interacting with users, likely due to their ubiquitous nature and the continuous data they provide, which are crucial for real-time recommendations and interventions. In contrast, PCs are less utilized for direct interaction in recommendation systems, which can be attributed to their relative immobility and lower user engagement compared to more integrated or personal devices. 
\FloatBarrier
\clearpage
\subsection{Algorithms used in recommender systems}

\begin{figure}[ht]
    \centering
    \includegraphics[width=0.49\textwidth]{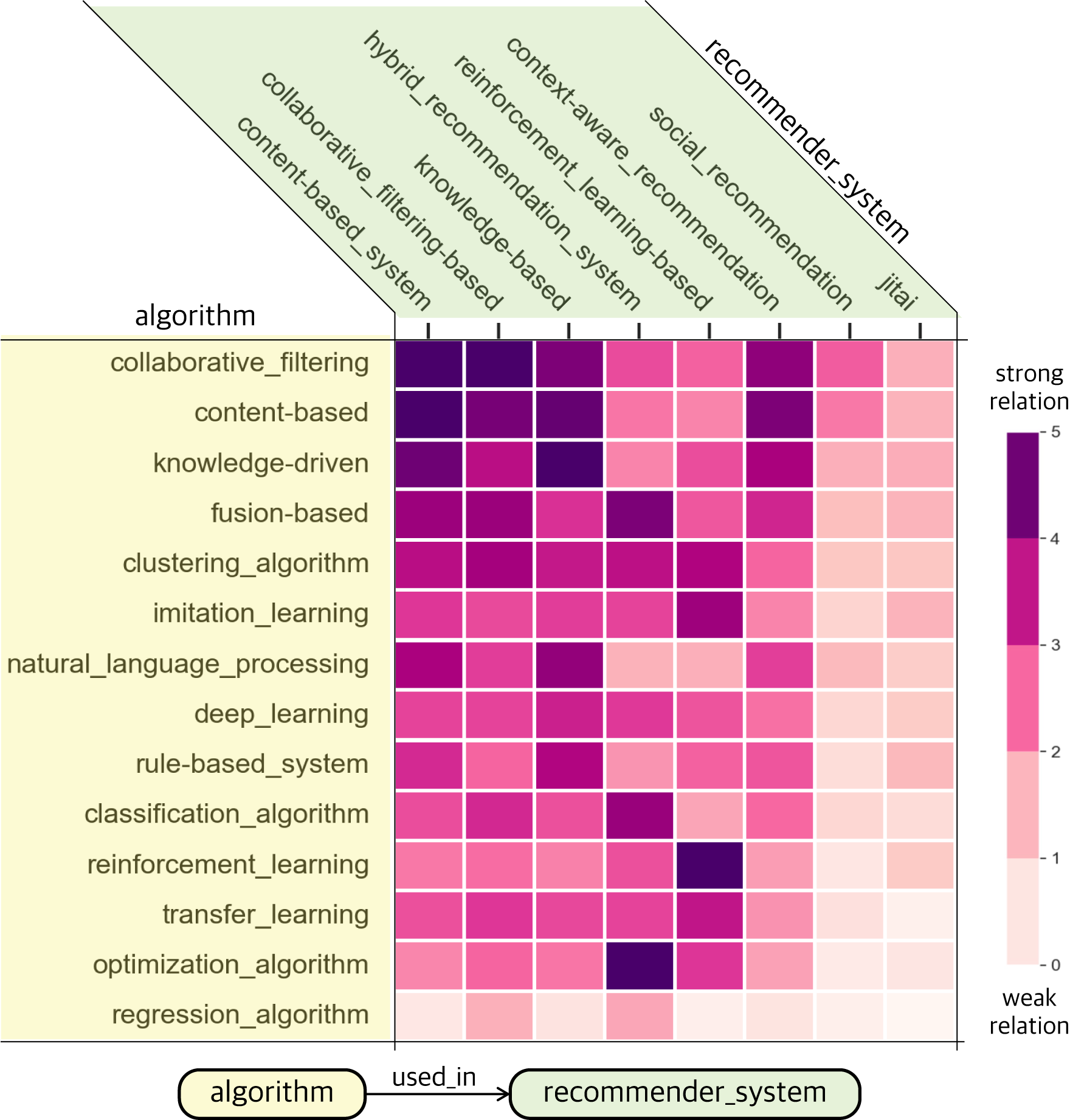}
    \caption{The relationships between algorithms (shown on the Y-axis with yellow highlight) and different types of recommender systems (shown on the X-axis with green highlight). Both of them are sorted based on the sum of the similarity per each row/column.}
    \label{fig:first_paper_result_3}
\end{figure}

The next comparison uses words related to the algorithm and the types of recommender systems, as shown in Figure \ref{fig:first_paper_result_3}. This heat map analysis reveals distinct patterns in the applicability of different algorithms across various types of recommender systems for human-building interaction and occupant support. Predominantly used algorithms include collaborative filtering, content-based, knowledge-driven, fusion-based, and clustering algorithms. NLP exhibits significant effectiveness in content-based and knowledge-driven systems due to its capacity to analyze and interpret complex user data and contextual information. In contrast, regression and optimization algorithms demonstrate a weaker association with most types of recommender systems examined. This may suggest that these more traditional algorithms are less favored for complex or dynamic recommendation tasks that require adaptation to new information or user behaviors. Moreover, algorithms such as RL display variable degrees of relationship strength across different types of recommender system, suggesting their applicative versatility and potential underexplored areas for their application. 
\FloatBarrier

\FloatBarrier

\section{Discussion}
\label{sec:discussion}
Given the prominence of recommender systems in health management and their growing relevance in the mobile health sector, substantial research has already been published in medical and mobile health journals. Therefore, the subsequent sections will focus on discussing the application of recommender systems in human-building interaction and context-aware support to elucidate their potential and identify gaps in this domain.  Specifically, this section will address the following three questions:

\begin{enumerate}
    \item What are the most commonly used algorithms?
    \item What are the most explored applications?
    \item What are the emerging application areas in which there are gaps?
    \item What are the benefits and limitations of text mining-based literature review?
\end{enumerate}

\begin{figure*}[h]
    \centering
    \includegraphics[width=1.0\textwidth]{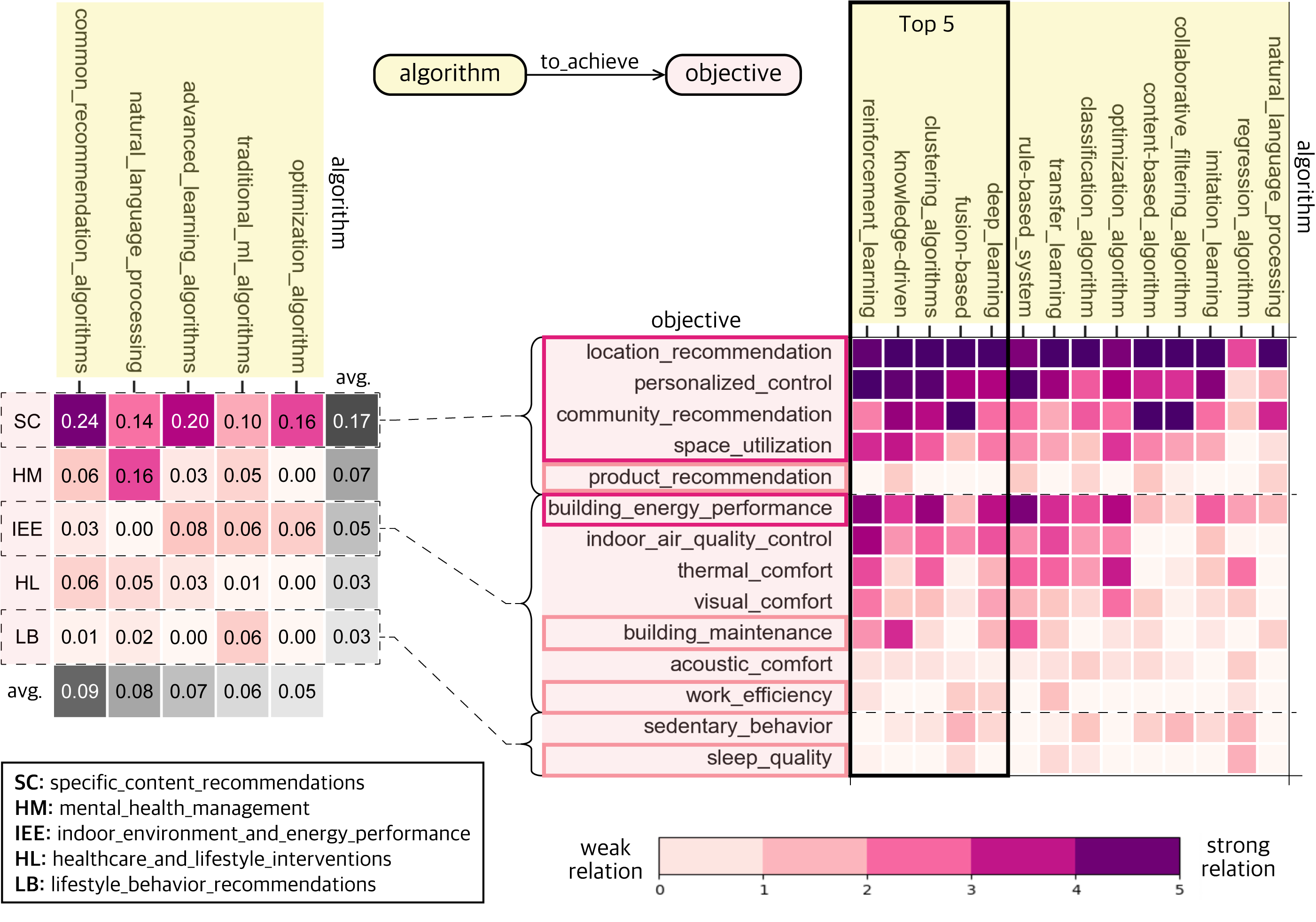}
    \caption{The figure on the left shows the relation between algorithms and recommendation objectives sorted based on usability. The figure on the right shows the relation between algorithms and human-building interaction applications sorted based on the group correlation and element correlation.}
    \label{fig:first_paper_discussion_1}
\end{figure*}

Figure \ref{fig:first_paper_discussion_1} presents the relationships between different types of algorithms and the various applications of recommender systems, with a specific emphasis on the application of each algorithm in the field of human-building interaction and occupant support. Both axes are organized based on the strength of relationships, with the horizontal axis representing the algorithms (arranged from left to right) and the vertical axis detailing the clusters of objectives (arranged from top to bottom). The following subsection will, based on the insights derived from this figure, elucidate key considerations that the research community can address in future studies.

\subsection{Algorithm selection for applications in human-building interaction and occupant support}
Based on the left side of Figure \ref{fig:first_paper_discussion_1}, it can be seen that the algorithms most commonly used for recommender systems include traditional methods such as content-based, collaborative filtering, and knowledge-based systems. However, for applications in human-building interaction and occupant support, where the variety of data types is extensive, the most commonly used recommendation algorithms are predominantly based on learning algorithms and traditional ML techniques. The five most commonly used recommendation algorithms, illustrated on the right side of Figure \ref{fig:first_paper_discussion_1} (highlighted with black borders), are RL, knowledge-based, clustering, fusion-based, and deep learning.

\textbf{Reinforcement learning-based} algorithm operates on the principle of learning optimal actions through interactions with the environment to maximize cumulative rewards \cite{712192}. These algorithms continually adjust their strategies based on user interaction feedback, refining their recommendations over time \cite{ALSAYED2024110085}. This adaptive capability makes RL-based algorithms particularly suitable for applications in the human-building interaction, where user preferences, environmental conditions, and energy consumption patterns frequently change \cite{ZHANG2019472}. In such applications, value-based methods like Deep Q-Network (DQN) and Double Deep Q-Network (DDQN) algorithms are extensively utilized for energy performance optimization \cite{9426901} and recommender system design \cite{CHEN2023110335}. Owing to their online learning capabilities, which include real-time strategy updates and experience replay mechanisms, these algorithms ensure that the models remain current and efficient \cite{WU2024109497}. Furthermore, RL-based algorithms are well-suited for addressing multi-objective optimization problems in complex environments. For instance, by integrating XGBoost with DQN, researchers have optimized indoor thermal comfort and energy consumption by predicting the impact of occupants' behaviors on indoor environments \cite{LIU2024111197}. Despite these advantages, RL algorithms face challenges such as the cold start problem, which is when the system lacks sufficient initial data to make accurate recommendations \cite{9001078}, time-intensive training processes \cite{WANG2020115036}, and poor interpretability \cite{WU2024109497}. Addressing the cold start issue requires substantial data for model training or the incorporation of prior knowledge during the design phase \cite{LIKA20142065, 10.1145/1352793.1352837}. Moreover, techniques like imitation learning and transfer learning are employed to reduce training times \cite{10.1145/3360322.3360849, 9043893}, while \cite{dai2023deciphering} employs explainable AI to address the challenge of poor interpretability.

\textbf{Knowledge-based} algorithms leverage domain-specific knowledge to generate recommendations, which is crucial when explicit user preferences are unavailable, and decisions must be guided by objective standards or expert knowledge \cite{Knowledge-Based}. In the context of human-building interaction, these algorithms can utilize detailed knowledge about building operations, environmental parameters, and user preferences to recommend optimal strategies \cite{RSforenergyefficiency, AnthonyJnr2021}. For example, a knowledge-based algorithm can be programmed to maintain acceptable indoor pollutant concentration levels. When sensors detect pollutant levels exceeding predefined thresholds, the system recommends increased ventilation to maintain air quality \cite{Felfernig2019}. Moreover, knowledge-based approaches are often used to enhance other algorithms. For instance, the HeartSteps algorithm, developed to promote physical activity, integrates RL with prior knowledge derived from historical data to overcome the cold start problem typical of traditional recommender systems \cite{liao2019personalizedheartstepsreinforcementlearning}.

\textbf{Clustering} algorithms enhance targeted and efficient recommendations by grouping similar items or users based on their attributes or behavioral characteristics using clustering techniques \cite{1427769}. Unlike collaborative filtering, which relies on user ratings or behavioral similarity \cite{10.1155/2009/421425}, clustering algorithms focus on identifying the inherent grouping structure within the data \cite{Ungar1998-da, Xiaojun2017}, thereby significantly mitigating the cold start problem commonly encountered in collaborative filtering due to data sparsity \cite{10.1007/11891321_4}. For instance, clustering methods like K-means \cite{764879} and DBSCAN \cite{10.1145/3068335} can perform extensive computations during the preprocessing stage, thereby reducing the need for real-time calculations and improving the operational efficiency of recommender systems. Despite their advantages in handling large-scale datasets, clustering algorithms exhibit limitations in dynamic environments, such as building control or indoor environment optimization \cite{Chen2007-bk}. Clustering algorithms typically require major computations, such as determining cluster centers and assigning data points to clusters, to be completed during the preprocessing phase \cite{1017616}. This means that whenever there are changes in the dataset, such as the addition of building operational data, the clustering model may need to be rerun to reflect these changes. This process is not automatic, contrasting with the real-time responsiveness of collaborative filtering \cite{10.1155/2009/421425}.

\textbf{Fusion-based (or Hybrid)} recommendation algorithms offer a solution to address the complexity of data and the diversity of requirements by integrating multiple data sources and recommendation techniques \cite{Burke2002}, such as collaborative filtering, content-based recommendation, and knowledge-driven approaches. These methods involve aggregating data representations from various sources prior to decision-making or combining recommendations from multiple sub-recommenders to construct an ensemble of recommendations \cite{Cano2017, 5712091}. For instance, \cite{AIELLO20184057} describes a decision support system that provides users with early warnings about disease transmission risks through a data aggregation system, which integrates indoor sensor data, mathematical models, and expert knowledge to generate informed recommendations. Similarly, \cite{VARLAMIS2022117775} proposed a smart data fusion method, based on the EM3 platform, which integrates sensor data with user habits and feedback to provide personalized energy-saving recommendations at optimal times. Additionally, \cite{CALVOBASCONES2024123834} presents a hybrid approach that combines ML algorithms with building geometry models to provide energy transition recommendations based on the compatibility of renewable energy sources within buildings. This integration allows the system to capitalize on the complementary strengths of diverse data and algorithms, thereby producing stronger, more comprehensive, and accurate recommendations \cite{10.1162/NECO_a_00079}. However, this approach also introduces challenges related to increased computational demands \cite{10.1145/3125620} and the complexity of algorithm tuning \cite{PRADHAN2020105784}.

\textbf{Deep learning-based} algorithms utilize complex neural network architectures to model user preferences and item characteristics within a recommendation framework \cite{LeCun2015}. These systems employ multi-layer nonlinear processing units to learn deep representations of features from complex user behaviors, item properties, and textual descriptions, enabling them to discern intricate patterns in large-scale data that traditional algorithms might overlook \cite{10.1145/3285029, Martins2020-sw}. For instance, in \cite{R2020113054}, deep learning models were used to address significant cold start issues in collaborative filtering systems by embedding information about users and their preferences directly into the model. Moreover, deep learning techniques are frequently applied to process unstructured data, such as text, images, and audio \cite{Betru2017-vh}. Such as Convolutional Neural Networks (CNNs) \cite{Girshick_2015_ICCV, He_2017_ICCV} are used for image recommendations, while Recurrent Neural Networks (RNNs) \cite{cho2014learningphraserepresentationsusing} handle sequential interaction data. Emerging technologies like attention mechanisms and Transformer models \cite{vaswani2023attentionneed} have been widely adopted in deep learning-based recommender systems \cite{10.1145/3285029}. These models focus more on the most pertinent parts of the input data, dynamically adjusting the importance weights of different segments to optimize information processing and feature extraction \cite{NIU202148}. For instance, in \cite{ZHANG2023109894}, the incorporation of time-aware self-attention mechanisms enhances the precision of capturing user preferences, significantly improving the performance of recommendation models.

\subsection{Well-established applications of recommender systems in human-building interaction}

From the specific perspective of applications using recommender systems shown on the right side of Figure \ref{fig:first_paper_discussion_1}, location-based recommendations, together with the corresponding optimization of space utilization, community recommendations, energy efficiency strategy recommendations, and personalized control recommendations aimed at optimizing indoor environments, are among the most popular (highlighted with red border). 

\textbf{Location recommendation} aims to suggest optimal places to users based on their preferences, behaviors, and contextual information \cite{10.1145/3154526}. These systems leverage various data sources, including geographic information, user location history, and social networks, to generate personalized location recommendations \cite{Rehman_Khalid_Madani_2017}. For instance, an enhanced recommendation method employs Item-based Collaborative Filtering to recommend new points of interest based on non-spatial attributes similar to previously liked locations \cite{6228105, 6427747}. However, traditional collaborative filtering approaches often face the challenge of high sparsity in the user-item matrix, which significantly limits their effectiveness \cite{10.1145/2629461}. To address this issue, researchers have proposed methods that integrate various Probability Matrix Factorization models to reduce dependency on specific data types \cite{7373327}. Additionally, the LCARS methodology proposed in \cite{10.1145/2629461} combines offline modeling and online recommendations not only to address the data sparsity issue but also to enhance explainability. In the context of human-building interaction, location recommendation systems are also frequently employed for internal space optimization to enhance energy efficiency, comfort, and functionality by optimizing the use of indoor spaces \cite{RSforenergyefficiency}. Given the complexity and diversity of data, methods based on deep learning and RL are widely applied in this domain \cite{9001078, 10.1145/3209219.3209244}. These algorithms often consider multidimensional factors such as users' occupancy patterns, environmental conditions, and personal preferences to provide effective solutions for indoor space management and optimization \cite{doi:10.1061/JCCEE5.CPENG-4973}.

\textbf{Personalized control recommendations} is another principal application. In the context of human-building interaction, these systems are commonly employed to optimize indoor conditions such as temperature, lighting, and air quality, involving the processing and analysis of time-series data on environmental parameters and user behaviors \cite{RSforenergyefficiency}. Traditional recommendation algorithms, like conventional collaborative filtering, typically lack the capability to handle time-series data and do not possess mechanisms for balancing multiple objectives in complex scenarios \cite{ISINKAYE2015261}. Consequently, personalized control recommendation systems for building energy patterns and indoor environment optimization frequently utilize RL, neural networks, and fusion algorithms to overcome the limitations of conventional recommendation algorithms and address the overload problem \cite{RSforenergyefficiency}. For example, \cite{LEI2022119742} proposed a deep RL-based occupant-centric HVAC control framework. This framework generates personalized HVAC control strategies for different users by analyzing environmental data and user feedback. Additionally, the ReViCEE algorithm has been developed, which analyzes individual and collaborative user preferences from historical data and indoor lighting conditions to generate recommendations aimed at stimulating user engagement and promoting behavioral changes towards energy conservation and sustainability \cite{KAR2019135}. Recent studies have further explored more personalized recommendations by integrating occupant physiological data with environmental data. For example, \cite{CHO2020107267} has explored calculating fatigue indices from activity data to provide lighting environments that alleviate fatigue. Another study proposed an emotion-oriented recommendation system for occupants, designed to suggest personalized indoor environmental quality settings based on emotional states \cite{KIM2024111396}.

\textbf{Building energy optimization} recommender systems enhance energy efficiency and sustainability by providing occupants with tailored energy management suggestions or appliance usage recommendations \cite{RSforenergyefficiency}. In the context of global energy challenges, these systems represent a significant future development trajectory. As shown in Figure \ref{fig:first_paper_discussion_1}, RL is the most commonly employed algorithm in this domain. RL demonstrates a unique advantage in its ability to adapt to dynamically changing environments by learning from user behavior and interactions with the surrounding environment \cite{Zhang_2022}. For instance, the RL-based system "recEnergy" leverages occupant feedback to retrain deep neural networks, thereby improving the effectiveness of energy-saving recommendations and increasing user acceptance rates \cite{9001078}. In addition, clustering algorithms, which generate energy setting recommendations based on similar cases \cite{8533391}, and knowledge-based algorithms, which produce energy-saving action suggestions by incorporating expert knowledge and predefined rules into energy management systems \cite{RSforenergyefficiency}, are also commonly employed due to their practicality. Furthermore, optimization algorithms, especially those integrated with building energy models, are widely used to address complex multi-objective optimization problems \cite{ROCHA2015203, CEBALLOSFUENTEALBA2019113953}. Recent research in this domain has focused on enhancing the temporal awareness of energy recommendation systems. This improvement aims to provide real-time, context-sensitive recommendations that are particularly valuable during emergencies, such as the COVID-19 pandemic \cite{RSforenergyefficiency}. 

\textbf{Community recommendation systems} are designed to recommend content, services, products, or other resources to community members (such as users or cohort) based on data related to the group's activities, interactions, and preferences \cite{terveen01beyond}. These systems have been widely implemented across various sectors, including community event recommendations, optimization of facility usage, facilitation of resident interactions, and emergency response and safety management \cite{smartcityreview1}. In the context of occupant comfort, \cite{QUINTANA2023109685} proposed cohort comfort models based on the preferences of similar users to recommend appropriate comfort settings tailored to specific user cohorts. For the algorithm inside, traditionally, these systems rely on established recommendation algorithms such as collaborative filtering and content-based filtering. Additionally, NLP techniques, such as topic modeling \cite{jelodar2019naturallanguageprocessinglda}, are frequently utilized to enhance the processing of unstructured data, which includes text and conversational data. Furthermore, advancements in LLMs, such as GPT (Generative Pre-trained Transformer) APIs, are being leveraged for complex semantic analysis \cite{bdcc8040036}. Ongoing research in this field aims to augment the contextual awareness of these algorithms while addressing challenges related to data privacy and security \cite{10.1145/3543846}. 

\subsection{Emerging applications and opportunities}

In Figure \ref{fig:first_paper_discussion_1}, the right side illustrates that certain recommendation objectives have notably low correlations with the underlying algorithmic concepts (highlighted with pink border). These weak links suggest that these areas may not be adequately addressed by current algorithmic frameworks, highlighting existing gaps and opportunities for innovation. These objectives include building fault detection and predictive maintenance, product recommendations, and specific environment optimization recommendations. 

\textbf{Automated fault detection and diagnostics (AFDD)} has rapidly evolved since the early 1990s \cite{Mah}. Traditionally, fault detection and diagnosis relied on rule-based algorithms and empirical formulas to generate reactive and preventive maintenance plans \cite{zhu2024surveypredictivemaintenancesystems}. With recent advances in AI, the focus has shifted towards ML-based systems \cite{8707108}. By analyzing sensor data and historical maintenance records, these systems can predict potential failures and recommend maintenance actions before issues escalate, enabling timely maintenance and minimizing downtime \cite{CHENG2020103087}. For instance, integrating ML algorithms with sensor data from HVAC systems allows for real-time insights into equipment health, thereby facilitating proactive maintenance measures \cite{SANZANA2022104445}. However, the current application of purely data-driven AFDD systems remains limited. This limitation arises from the difficulty and high cost of acquiring fault data. Moreover, in real-world FDD applications, it is possible to encounter unknown system faults that were not accounted for during the training phase \cite{CHEN2022112395}. The lack of comprehensive training data prevents these systems from covering all building and system operating conditions effectively. Recently, the advancements in generative AI may address the issue of data scarcity \cite{BAASCH2021100087}. Additionally, the adoption of Non-Intrusive Load Monitoring (NILM) as a replacement for sub-metering significantly reduces the cost and improves the accuracy of energy consumption data acquisition, particularly at the appliance level \cite{RSforenergyefficiency}. These developments have the potential to substantially lower the cost and complexity of developing anomaly detection systems.

\textbf{Product recommendation systems} are extensively utilized in e-commerce and retail sectors to enhance user experiences by providing personalized product suggestions based on user behavioral data or basic information such as purchase history, browsing records, and user reviews \cite{Chen2015, 011f100966f4451f9c59bade4700434c}. Recently, studies have highlighted that integrating environmental data with user preferences in product recommendation systems can significantly improve the relevance and effectiveness of recommendations. For instance, \cite{8071982} designed a recommendation system based on a sensor network that extracts user preferences to suggest indoor comfort products, resulting in a 16\% increase in consumer purchases. Additionally, integrating appliance energy consumption data into recommendation systems can guide consumers toward purchasing energy-efficient products, thereby contributing to improved building energy efficiency \cite{RSforenergyefficiency}. For example, a study proposed a recommendation system leveraging NILM techniques to detect appliance usage and provide smart grid users with suggestions for energy-efficient appliances that are most relevant or beneficial to them \cite{https://doi.org/10.1049/iet-gtd.2016.1615}.
Additionally, with the deregulation of the retail electricity market, research has also employed recommender systems to provide electricity customers with personalized recommendations for electricity providers and electricity plans tailored to their individual consumption needs, aiming to help them reduce energy costs \cite{ZHAO2021117191}. Furthermore, product recommendation systems have also been applied to suggest building design strategies \cite{BORATTO201779} and energy transition solutions \cite{CALVOBASCONES2024123834}, offering decision support to occupants.

 \textbf{Specific environment optimization} through recommender systems also represents one of the key future direction for development. With the advancement of thermal comfort models, recent research has shifted towards offering optimization recommendations for specific settings based on their unique features, such as enhancing sleep quality and work efficiency in sleeping and working environments \cite{s23156670, 8615217}. These specialized environments require not only environmental data but also an integration of interdisciplinary domain knowledge and individual physiological data or real-time feedback to furnish more accurate recommendations \cite{Pulantara2018-pg}. With the widespread adoption of smart wearable devices and smart home technologies, the real-time acquisition of such data has become feasible. For instance, the Cozie app \cite{Tartarini_2023}, based on the Apple Watch and Fitbit, can be utilized to gather micro-ecological momentary assessment data and physiological data. By integrating occupant feedback with other sensor data, personalized adjustment recommendations can be created \cite{10.1145/3563357.3566135}. However, such highly personalized systems pose challenges in the initial data acquisition, processing, and storage of large volumes of sensitive data \cite{RSforenergyefficiency}. 

\subsection{Implications and future directions of text mining-based literature review}

Text mining has become a powerful tool for conducting literature reviews, offering notable advantages that enhance both efficiency and comprehensiveness. Its ability to process large-scale datasets efficiently allows researchers to analyze vast amounts of literature that would be unmanageable with traditional manual methods \cite{Mah}. This scalability enables the identification of hidden patterns and trends across studies that might otherwise go unnoticed. Additionally, text mining reduces inherent human bias in manual reviews, supporting a more objective synthesis of findings.

Despite its strengths, text mining has limitations. The effectiveness of these techniques heavily depends on the quality and structure of the input data. Noise from poorly managed datasets or insufficient data pre-processing can lead to incomplete or misleading results \cite{Bauer2021-mn}. Moreover, some text mining methods face challenges in interpretability, making it difficult for researchers to fully understand how specific conclusions are reached, which can undermine the credibility of the findings \cite{doi:10.1177/10944281221124947}. Text mining may also overlook nuanced or contextual information, such as subtle implications in methodology or theoretical frameworks, which human reviewers are better equipped to identify\cite{Mah}.

Looking forward, several strategies can enhance the integration of text mining in literature reviews. A promising approach involves combining text mining techniques with traditional review methods, creating a hybrid framework that balances automation with the depth of human expertise. In addition, integrating transformer-based pertrained models, such as BERT \cite{devlin2019bertpretrainingdeepbidirectional} and LLMs, can further improve the understanding of complex academic texts \cite{ZHU2023106876}. Fine-tuning these models on domain-specific datasets could significantly enhance the precision of extracting key information, revolutionizing automated literature reviews \cite{ANISUZZAMAN2024}. Finally, developing standardized frameworks for applying text mining in literature reviews can improve reproducibility and comparability across studies, fostering greater consistency across disciplines.

\section{Conclusion}
\label{sec:conclusion}

This study employs text mining techniques to perform a comprehensive review of the literature in the context of human-building interaction and occupant context-aware support, and recommendation systems. By extracting and analyzing trends, patterns, and relationships from 27,595 articles sourced from Elsevier journals, findings that are typically elusive to traditional review methods are revealed. Compared to other studies, this study provides the first comprehensive analysis of the applications of recommendation systems in the context of human-building interaction, including energy efficiency, occupant health, and indoor environmental quality. It also quantifies the interrelationships among different components (data sources, algorithm types, system types, platforms, and recommendation objectives) of recommendation systems. 

The results indicate that recommendation systems have been widely applied to specific content, such as space optimization, location recommendations, and personalized control suggestions. However, the application of recommendation system methodologies to optimize indoor environments and energy efficiency is not as common. This discrepancy is likely due to the fact that these applications often require the integration of interdisciplinary domain knowledge and extensive sensor data. Moreover, while various data sources related to user context, activities, and real-time feedback are frequently utilized, other critical data sources, including those related to physiological and indoor environmental factors, are underutilized despite their importance for optimizing specific environments or objectives. In addition, traditional recommendation algorithms such as collaborative filtering, content-based, and knowledge-based methods are widely used in this area. Lastly, due to the robust data collection capabilities of wearable devices and their user-friendly interfaces, they are commonly utilized as platforms to implement recommendation systems.

In addition, this work addresses several key questions regarding the application of recommender systems in human-building interaction and occupant context-aware support. Specifically, it examines the most commonly utilized algorithms, the most explored applications, and emerging opportunities in this field. Among these, deep reinforcement learning stands out as the predominant algorithm for building-related applications, primarily due to its ability to handle complex inputs and self-update based on occupant feedback. Although certain applications such as location recommendation, personalized control, and energy optimization of buildings are widely adopted, significant opportunities for innovation remain in emerging areas. These include areas such as predictive maintenance of building systems, recommendations for building design, appliances, and electricity plans, as well as specific environment optimization, such as those designed to improve sleep and work productivity based on occupant feedback.

This review also acknowledges some limitations. The methodology may not fully comprehend the nuances and contextual subtleties of academic texts, potentially overlooking the impact of pioneering research. Furthermore, NLP systems might misinterpret terms with multiple meanings or highly specialized terminology without sufficient contextual data. Future directions include integrating transformer-based pertrained models (such as BERT) and LLM, which could improve the comprehension of complex academic texts. Furthermore, fine-tuning these models on specific academic datasets could significantly improve the precision with which key information is extracted, potentially transforming the landscape of automated literature reviews.

\section*{Reproducibility}

This analysis can be reproduced using the data and code found in this GitHub repository: \url{https://github.com/buds-lab/recommender-sys-for-buildings-textmining-review}.

\section*{CRediT authorship contribution statement}
\textbf{Wenhao Zhang:} Writing – original draft, Visualization, Software, Methodology, Investigation, Formal analysis, Data curation, Conceptualization. \textbf{Matias Quintana:} Writing – review \& editing, Conceptualization. \textbf{Clayton Miller:} Writing – review \& editing, Supervision, Resources, Project administration, Funding acquisition, Conceptualization.

\section*{Funding}
The research was supported by a Ph.D. scholarship from the Singapore Ministry of Education (MOE) and by the National Research Foundation, Prime Minister’s Office, Singapore, under its Campus for Research Excellence and Technological Enterprise (CREATE) program.

\section*{Acknowledgements}
The authors thank Dr. Mahmoud Abdelrahman for his guidance on using the Elsevier API for article retrieval.

\section*{Declaration of generative AI and AI-assisted technologies in the writing process}
During the preparation of this work, the authors used ChatGPT to perform grammar checking and language polishing. After using this tool/service, the authors reviewed and edited the content as needed and take full responsibility for the content of the published article.

\bibliographystyle{unsrt}

\end{document}